\documentstyle[psfig]{mn}

\newif\ifAMStwofonts
\AMStwofontstrue

\newcommand{\ea}{{\it et~al.}}
\newcommand{\dg}{\nobreak^\circ}

\newcommand{\ho}{\mbox{$\mbox{H}_0$} }
\newcommand{\simlt}{\mbox{$\stackrel{<}{_{\sim}}$} }

\newcommand{\kmspmpc}{\,\hbox{km}\,\hbox{s}^{-1}\,\hbox{Mpc}^{-1}}

\newcommand{\eg}{{\it e.g.\ }}

\newcommand{\uk}{$\mu$K}
\newcommand{\be}{\begin{equation}}
\newcommand{\ee}{\end{equation}}

\ifoldfss
  \ifCUPmtlplainloaded \else
    \NewTextAlphabet{textbfit} {cmbxti10} {}
    \NewTextAlphabet{textbfss} {cmssbx10} {}
    \NewMathAlphabet{mathbfit} {cmbxti10} {} 
    \NewMathAlphabet{mathbfss} {cmssbx10} {} 
  \fi
  \ifAMStwofonts
    \ifCUPmtlplainloaded \else
      \NewSymbolFont{upmath} {eurm10}
      \NewSymbolFont{AMSa} {msam10}
      \NewMathSymbol{\upi}     {0}{upmath}{19}
      \NewMathSymbol{\umu}     {0}{upmath}{16}
      \NewMathSymbol{\upartial}{0}{upmath}{40}
      \NewMathSymbol{\leqslant}{3}{AMSa}{36}
      \NewMathSymbol{\geqslant}{3}{AMSa}{3E}

       \let\le=\leqslant
       \let\ge=\geqslant
    \fi
  \fi
\fi 

\ifnfssone
  \newmathalphabet{\mathit}
  \addtoversion{normal}{\mathit}{cmr}{m}{it}
  \addtoversion{bold}{\mathit}{cmr}{bx}{it}
  \newmathalphabet{\mathbfit} 
  \addtoversion{normal}{\mathbfit}{cmr}{bx}{it}
  \addtoversion{bold}{\mathbfit}{cmr}{bx}{it}
  \newmathalphabet{\mathbfss} 
  \addtoversion{normal}{\mathbfss}{cmss}{bx}{n}
  \addtoversion{bold}{\mathbfss}{cmss}{bx}{n}
  \ifAMStwofonts
    \ifCUPmtlplainloaded \else
      \UseAMStwoboldmath
      \makeatletter
      \new@mathgroup\upmath@group
      \define@mathgroup\mv@normal\upmath@group{eur}{m}{n}
      \define@mathgroup\mv@bold\upmath@group{eur}{b}{n}
      \edef\UPM{\hexnumber\upmath@group}
      \new@mathgroup\amsa@group
      \define@mathgroup\mv@normal\amsa@group{msa}{m}{n}
      \define@mathgroup\mv@bold\amsa@group{msa}{m}{n}
      \edef\AMSa{\hexnumber\amsa@group}
      \makeatother
      \mathchardef\upi="0\UPM19
      \mathchardef\umu="0\UPM16
      \mathchardef\upartial="0\UPM40
      \mathchardef\leqslant="3\AMSa36
      \mathchardef\geqslant="3\AMSa3E

       \let\le=\leqslant
       \let\ge=\geqslant
    \fi
  \fi
\fi 

\ifnfsstwo
  \DeclareMathAlphabet{\mathbfit}{OT1}{cmr}{bx}{it}
  \SetMathAlphabet\mathbfit{bold}{OT1}{cmr}{bx}{it}
  \DeclareMathAlphabet{\mathbfss}{OT1}{cmss}{bx}{n}
  \SetMathAlphabet\mathbfss{bold}{OT1}{cmss}{bx}{n}
  \ifAMStwofonts
    \ifCUPmtlplainloaded \else
      \DeclareSymbolFont{UPM}{U}{eur}{m}{n}
      \SetSymbolFont{UPM}{bold}{U}{eur}{b}{n}
      \DeclareSymbolFont{AMSa}{U}{msa}{m}{n}
      \DeclareMathSymbol{\upi}{0}{UPM}{"19}
      \DeclareMathSymbol{\umu}{0}{UPM}{"16}
      \DeclareMathSymbol{\upartial}{0}{UPM}{"40}
      \DeclareMathSymbol{\leqslant}{3}{AMSa}{"36}
      \DeclareMathSymbol{\geqslant}{3}{AMSa}{"3E}

       \let\le=\leqslant
       \let\ge=\geqslant
    \fi
  \fi
\fi 

\ifCUPmtlplainloaded \else
  \ifAMStwofonts \else 
    \def\upi{\pi}
    \def\umu{\mu}
    \def\upartial{\partial}
  \fi
\fi

\title[Constraints on cosmological parameters]
{Constraints on cosmological parameters from recent measurements of CMB anisotropy}
\author[S. Hancock et al.]
       {S.~Hancock,$^1$ G.~Rocha,$^{1,3}$ A.N. Lasenby$^1$ and
C.M. Guti\'{e}rrez$^2$\\
	$^1$Mullard Radio Astronomy Observatory, Cavendish Laboratory,
Madingley Road,  Cambridge CB3 OHE, UK\\
	$^2$Instituto de Astrofisica de Canarias, 38200 La Laguna,
Tenerife, Spain\\
	$^3$Department of Physics, Kansas State University, Manhattan,
 KS 66506, USA}

\date{Accepted .
      Received ;
      in original form }

\pagerange{\pageref{firstpage}--\pageref{lastpage}}
\pubyear{1996}

\begin{document}

\maketitle

\label{firstpage}

\begin{abstract}
A key prediction of cosmological theories for the origin and evolution of
structure in the Universe is the existence of a `Doppler peak' in the angular
power spectrum of cosmic microwave background (CMB) fluctuations.  We present
new results from a study of recent CMB observations which provide the first
strong evidence for the existence of a `Doppler Peak' localised in both angular
scale and amplitude. This first estimate of the angular position of the peak is
used to place a new direct limit on the curvature of the Universe,
corresponding to a density of $\Omega=0.7^{+0.8}_{-0.5}$, consistent with a
flat Universe.  Very low density `open' Universe models are inconsistent with
this limit unless there is a significant contribution from a cosmological
constant. For a flat standard Cold Dark Matter dominated Universe we use our
results in conjunction with Big Bang nucleosynthesis constraints to determine
the value of the Hubble constant as $\ho=30-70\kmspmpc$ for baryon fractions
$\Omega_b=0.05$ to $0.2$. For $\ho=50\kmspmpc$ we find the primordial spectral
index of the fluctuations to be $n=1.1 \pm 0.1$, in close agreement with the
inflationary prediction of $n \simeq 1.0$.  
\end{abstract}

\begin{keywords}
cosmology -- cosmic microwave background.
\end{keywords}

\section{Introduction}

Observations of the Cosmic Microwave Background (CMB) radiation provide
information about epochs and physical scales that are inaccessible to
conventional astronomy. In contrast to traditional methods of determining
cosmological parameters, which rely on the combination of results from local
observations \cite{os}, CMB observations provide direct measurements
\cite{be87,white} over cosmological scales, thereby avoiding the systematic
uncertainties and biases associated with conventional techniques. The principal
cosmological information is contained in the acoustic peaks
\cite{be87,hu,scott} in the power spectrum, which are generated during acoustic
oscillations of the photon-baryon fluid at recombination \cite{cdm2}.  The main
acoustic peak, sometimes referred to as `the first Doppler peak', is a strong
prediction of contemporary cosmological models with adiabatic fluctuations and
is expected to occur on an angular scale $\sim 1\dg$. (In topological defect
theories of structure formation, the first Doppler peak is expected to occur on
smaller angular scales (e.g. Magueijo \ea\ 1996) or be of much smaller amplitude
\cite{pen} than in inflationary theories. This is discussed further below.)
The observation of this peak is thus a major goal of observational
cosmology. In the case that it is not observed, this could imply either that
medium-scale primordial CMB fluctuations had been wiped out by reionization
\cite{cdm2}, or perhaps that there is a fundamental flaw in our theory. On the
contrary, a conclusive observation of the first peak would provide strong
support for current theoretical models and the determination of its angular
position would constitute a direct probe of the large scale geometry of the
Universe.  The angular scale $l_p$ of the main peak reflects the size of the
horizon at last scattering of the CMB photons and thus depends almost entirely
\cite{hu,kamio} on the total density of the Universe according to $l_p \propto
1/\sqrt{\Omega}$.  In conventional inflationary theory \cite{guth}, one expects
the Universe to be flat with $\Omega=1.0$, which can be achieved if the total
mass density is equivalent to the critical density or if there is a
contribution from a cosmological constant $\Lambda$.  The height of the peak
provides additional cosmological information since it is directly proportional
to the fractional mass in baryons $\Omega_b$ and also varies according to the
expansion rate of the Universe as specified by the Hubble constant
$\mbox{H}_0$; in general \cite{hu} for baryon fractions $\Omega_b \simlt 0.05$,
increasing \ho reduces the peak height whilst the converse is true at higher
baryon densities.  Furthermore, by measuring the amplitude of the intermediate
scale CMB fluctuations relative to those on large scales it is possible to
place tight limits on the spectral slope $n$ of the initial primordial spectrum
of fluctuations. The latter is predicted by inflationary theory to be
approximately scale invariant, in which case $n \simeq 1.0$, although (in particular
versions of inflationary theory) the presence of a background of primordial
gravity waves would lead to lower values of $n$, via the relation $C_2^T/C_2^S
\approx 7(1-n)$, where $C_2^T/C_2^S$ is the ratio of tensor to scalar
contributions to the quadrupole component of the CMB power spectrum
\cite{crittenden,steinhardt}. (Linkages between parameters in more general
theories of inflation are discussed in Liddle (1997).) Thus, in summary, by
comparing large and intermediate scale CMB observations and tracing out the
Doppler peak, it is possible to directly estimate $\Omega$, $\Omega_b$ and \ho
and to probe inflationary theory and the existence of primordial gravity
waves. Recent improvements in the quality of CMB data, in particular on the
angular scales probed by the CAT and Saskatoon experiments, now make this
exercise of great interest.

\section{Method}

Clear detections of CMB anisotropy have now been reported by a number of
different groups, including the COBE satellite
\cite{smoot,bennett}; ground-based switching experiments such as Tenerife
\cite{me96,nature94}, Python \cite{python}, South Pole \cite{spole} and
Saskatoon \cite{sask}; balloon mounted instruments such as ARGO \cite{argo},
MAX \cite{max}, and MSAM \cite{msam1,msam2} and more recently the ground-based
interferometer CAT \cite{cat}.  Given the difficulties inherent in observing
CMB anisotropy, it is possible that some of these results are contaminated by
foreground effects and it is clear that determining the form of the CMB power
spectrum in order to trace out the Doppler peak requires a careful, in-depth
consideration of the CMB measurements from the different experiments within a
common framework. The full details including a discussion of foreground
contamination are presented in Rocha \ea\ \shortcite{gr} and here we
present our principal findings.  We consider all of the latest CMB
measurements, including new results from COBE, Tenerife, MAX, Saskatoon and
CAT, with the exception of the MSAM results (see below) and the MAX detection
in the Mu Pegasi region which is contaminated by dust emission \cite{fischer}.

\begin{figure}
\centerline{\psfig{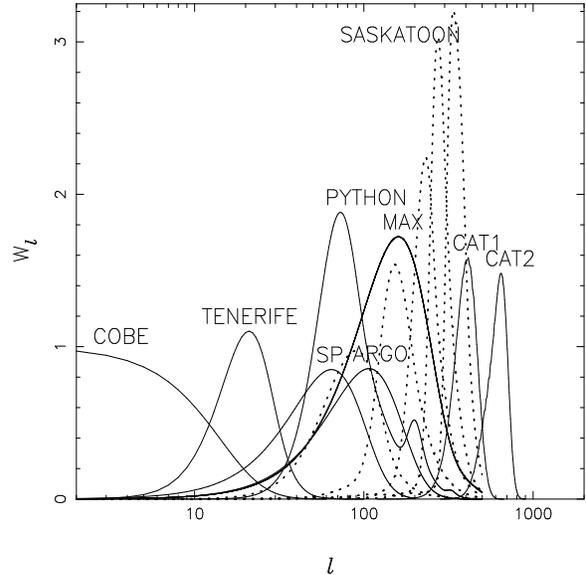}}
\caption{The window functions for the experiments listed in Table~1 \label{fig:windows}}
\end{figure}

The competing models for the origin and evolution of structure predict
\cite{be87,hu}, the shape and amplitude of the CMB power spectrum and its
Fourier equivalent, the autocorrelation function $C(\theta) =<\Delta T({\bf
n_1})\Delta T({\bf n_2})>$ where ${\bf n_1} \cdot {\bf n_2} =\cos \theta$.
Expanding the intrinsic angular correlation function $C(\theta)$ in terms of
spherical harmonics one obtains
\be
C(\theta)= \sum_{l\ge2}^{\infty} (2l+1) C_l
P_l(\cos\theta) /4 \pi,
\ee
where low order multipoles $l$ correspond to large angular scales $\theta$ and
large $l$-modes are equivalent to small angles on the sky.  The $C_l$'s are
predicted by the cosmological theories and contain all of the relevant
statistical information for models described by Gaussian random fields
\cite{be87}. The different experiments sample different angular scales
according to their {\em window functions} $W_l$
\cite{whitewindow1,whitewindow2}. 
The window function $W_l$ specifies the relative sensitivity of an
experiment to a given $l$-mode, and the observed power in CMB fluctuations as
seen through a window $W_l$ is given by
\be C_{obs}(0)=\left (\frac{\Delta
T_{obs}}{T} \right )^2 = \sum_{l \ge 2}^{ \infty} (2l+1) C_l W_l/4 \pi.
\label{eq:cobs}
\ee 
Given $W_l$, then for the $C_l$'s corresponding to the theoretical model under
consideration it is possible to obtain the value of $\Delta T_{obs}$ one would
expect to observe using the chosen experiment.  This value can then be compared
to the value actually observed to test the cosmological model.  Shown in
Fig.~\ref{fig:windows} are the window functions 
for the various configurations of the experiments considered.

On the largest scales corresponding to small $l$, new COBE \cite{bennett} and
Tenerife \cite{me96} results improve the power spectrum normalisation, whilst
significant gains in knowledge at high $l$ are provided by new results from the
Saskatoon and CAT experiments. The full data set spans a range of $2$ to $\sim
700$ in $l$, sufficient to test for the main Doppler peak out to $\Omega =0.1$.
We take the reported CMB detections and convert them to a common framework of
flat bandpower results \cite{bond1,bond2} as given in Table 1. This is carried
out as follows.
\begin{table*}
\begin{minipage}{4.5in}
\caption{Details of data results used}
\begin{tabular}{|l|c|c|c|c|c|c|} \hline
Experiment & $\Delta T_l$ ($\mu$K) & $\sigma$ ($\mu$K) & $l_e$ & $l_l$  &$l_u$
& Reference \\ 
\hspace*{.2in} \\
 COBE       &     27.9 &     2.5 &   6 &   2 &   12 & \cite{bennett}\\
 Tenerife   &     34.1 &     12.5 &   20 &   13 &   31 & \cite{me96}\\
 PYTHON     &     57.2 &     16.4 &   91 &   50 &   107 & \cite{python}\\
 South Pole &     39.5 &     11.4 &   57 &   31 &   106 & \cite{spole}\\
 ARGO       &     39.1 &     8.7 &   95 &   52 &   176 & \cite{argo}\\
 MAX GUM    &     54.5 &     13.6 &   145 &   78 &   263 & \cite{max}\\
 MAX ID     &     46.3 &     17.7 &   145 &   78 &   263 & ``\\
 MAX SH     &     49.1 &     19.1 &   145 &   78 &   263 & ``\\
 MAX PH     &     51.8 &     15.0  &   145 &   78 &   263 & ``\\
 MAX HR     &     32.7 &     9.5 &   145 &   78 &   263 & ``\\
 Saskatoon1 &     49.0 &     6.5 &   86 &   53 &   132  & \cite{sask}\\
 Saskatoon2 &     69.0 &     6.5 &   166 &   119 &   206 & ``\\
 Saskatoon3 &     85.0 &     8.9 &   236 &   190 &   274 & ``\\
 Saskatoon4 &     86.0 &     11.0 &   285 &   243 &   320 & ``\\
 Saskatoon5 &     69.0 &     23.5 &   348 &   304 &   401 & ``\\
 CAT1       &     50.8 &     15.4 &   396 &   339 &   483 & \cite{cat}\\
 CAT2       &     49.0 &     16.9 &   608 &   546 &   722 & ``\\
\hline
\end{tabular}
\end{minipage}
\end{table*}

The CMB anisotropy measurements are converted to bandpower estimates $\Delta
T_l \pm \sigma$ assuming in each case a flat spectrum of $C_l$ centred on the
effective multipole $l_e$ (see below) of the window function. $l_l$ and $l_u$
represent the lower and upper points at which the window of each configuration
reaches half of its peak value.  In order to use the observed anisotropy levels
to place constraints on the CMB power spectrum one must in general know the
form of the $C_l$ under test.  However, in most cases the form of $C_l$ can be
represented by a flat spectrum $C_l \propto C_2/(l(l+1))$ over the width of a
given experimental window, so that the bandpower is $\Delta
T_l/T=\sqrt{C_{obs}(0)/I(W_l)}$, where we define $I(W_l)$ according to Bond
\shortcite{bond1,bond2} as $I(W_l)=\sum_{l=2}^{\infty}
(l+0.5)W_l/(l(l+1)$. This bandpower estimate is centred on the effective
multipole $l_e=I(lW_l)/I(W_l)$.  In many instances experimenters now report
results directly for a flat spectrum and when this is not so we have converted
the quoted power in fluctuations into the equivalent flat band estimate.  Each
group has obtained limits on the intrinsic anisotropy level using a likelihood
analysis (see \eg Hancock \ea\ 1994), which incorporates
uncertainties due to random errors, sampling variance \cite{samvariance} and
cosmic variance \cite{cosvariance2,cosvariance1}.  The errors in $\Delta T_l$
quoted in column 3 of Table~1 are at 68 \% confidence and have been obtained by
averaging the difference in the reported 68\% upper and lower limits and the
best fit $\Delta T_l$. Since the form of the likelihood function is in general
only an approximation to a Gaussian distribution this averaging introduces a
small bias into the results \cite{gr}. With the exception of Saskatoon the
errors include uncertainties in the overall calibration.  There is a $\pm 14$\%
calibration error in the Saskatoon data, but since the Saskatoon points are not
independent this will apply equally to all five points \cite{sask}. We discuss
below how this is included in the analysis.  It is not possible to ascribe an
error to each experiment to represent the likely degree of residual Galactic
contamination present in its results, since this is not known at
present. However, as emphasized above, if there is any evidence that the degree
of contamination in an experimental point could be significant, we have not used that
point.

Results from the MSAM experiment are not included here, because they do not
provide an independent measure of the power spectrum since their angular
sensitivity and sky coverage are already incorporated within the Saskatoon
measurements. Netterfield \ea\ \shortcite{sask} report good agreement between
the MSAM double difference results and Saskatoon measurements, although the
discrepancy with the MSAM single difference data is yet to be resolved. The
window functions for the COBE and Tenerife experiments are independent at the
half-power points, thus justifying their joint use even though their sky areas
overlap.

\begin{figure}
\centerline{\psfig{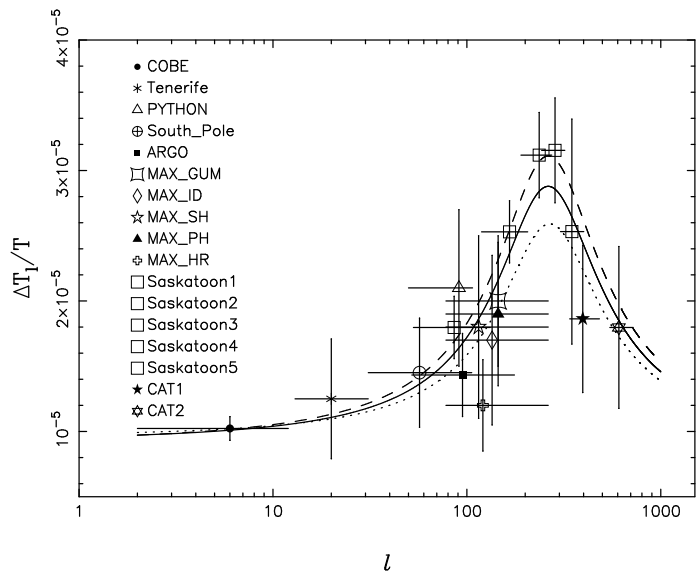}}
\caption{The data points from Table 1 are shown compared to the best fit
analytical CDM model. The dotted and dashed lines show the best fit models
which are obtained when the Saskatoon calibration is adjusted by $\pm 14\%$. The
data points from the MAX experiment are shown offset in $l$ for clarity \label{fig:fig2a}}
\end{figure}

The data points from Table 1 are plotted in Figure~\ref{fig:fig2a},
in which the horizontal bars represent the range of $l$ contributing to each
data point. There is a noticeable rise in the observed power spectrum at
$l\simeq 200$, followed by a fall at higher $l$, tracing out a clearly defined
peak in the spectrum.  In the past several groups \cite{scott,kami,rhatra} have
attempted to determine the presence of a Doppler peak, but only now are the
data sufficient to make a first detection and to put constraints on the closure
parameter $\Omega$.  As a first step, we adopt a simple three parameter model
of the power spectrum, which we find adequately accounts
for the properties of the principal Doppler peak for both standard Cold Dark
Matter (CDM)  models \cite{cdm1,cdm2} and open Universe ($\Omega<1$) models
\cite{kami}. The functional form chosen is a modified version of that used in 
Scott, Silk \& White \shortcite{scott} --- we choose the following: 
\be
l(l+1)C_l=6C_2\left(1+\frac{A_{peak}}{1+y(l)^2}\right ) {\huge /}
\left(1+\frac{A_{peak}}{1+y(2)^2}\right )
\label{eq:peak}
\ee 
where $y(l)=(\log_{10}l-log_{10}(220/\sqrt{\Omega}))/0.266$. In this
representation $C_2$ specifies the power spectrum normalisation, whilst the
first Doppler peak has height $A_{peak}$ above $C_2$, width $\log_{10}l=0.266$
and for $\Omega=1.0$ is centred at $l\simeq 220$.  By appropriately specifying
the parameters $C_2$, $A_{peak}$ and $\Omega$ it is possible to reproduce to a
good approximation the $C_l$ spectra corresponding to standard models of
structure formation with different values of $\Omega$, $\Omega_b$ and
$\mbox{H}_0$. Such a form will not reproduce the structure of the {\em
secondary\/} Doppler peaks, but we have checked the model against the overall
form of the $\Omega =1$ models of Efstathiou and the open models reported in
Kamionkowsky \ea\ \shortcite{kami} and find that this form adequately reflects
the properties of the main peak. This satisfies our present considerations
since the current CMB data are not yet up to the task of discriminating the
secondary peaks.  Varying the three model parameters in
equation~(\ref{eq:peak}) we form $C_l$ spectra corresponding to a range of
cosmological models, which are then used in equation~(\ref{eq:cobs}) to obtain
a simulated observation for the $i$th experiment, before converting to the
bandpower equivalent result $\Delta T_l[C_2,A_{peak},\Omega](i)$.  The
chi-squared for this set of parameters is given by
\begin{displaymath}
\chi^2(C_2,A_{peak},\Omega)=\sum_{i=1}^{nd}
\frac{(\Delta T_l^{obs}(i) -\Delta T_l[C_2,A_{peak},
\Omega](i))^2}{\sigma_i^2}, 
\label{eq:chi}
\end{displaymath}
where $nd$ is the number of data points in Table~1 and the relative likelihood function is
formed according to $L(C_2,A_{peak},\Omega) \propto \exp( -
\chi^2(C_2,A_{peak},\Omega)/2)$.  We vary the power spectrum normalisation
$C_2$ within the 95 \% limits for the COBE 4-year data \cite{bennett} and
consider $A_{peak}$ in the range 0 to 30 and values of the density parameter up
to $\Omega=5$.  The data included in the fit are those from Table 1. Because of
the $\pm 14\%$ calibration error in the Saskatoon data, the
likelihood function is evaluated for three cases: (i) that the
calibration is correct, (ii) the calibration is the lowest allowed value and
(iii) the calibration is the maximum allowed value.  In each case the likelihood
function is marginalised over $C_2$ before calculating limits on the remaining
two parameters according to Bayesian integration with a uniform prior.

\section{Results and Discussion}

\begin{figure}
\centerline{\psfig{figure=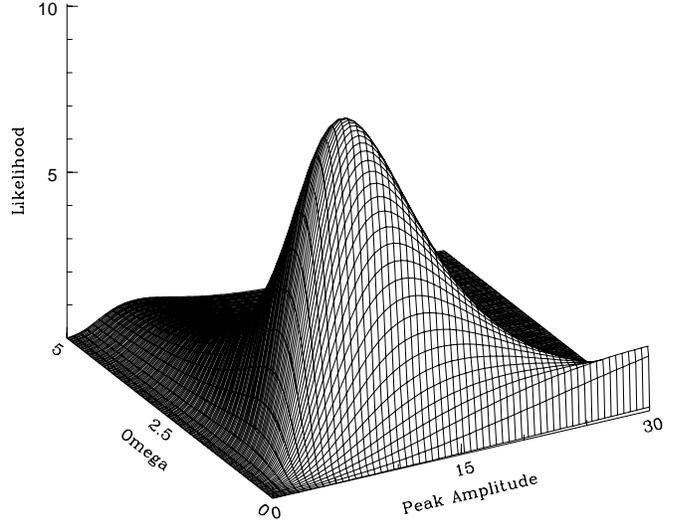,width=4.0in,angle=0}}
\caption{The likelihood surface for $\Omega$ and $A_{peak}$. (The nominal
Saskatoon calibration is assumed.) \label{fig:3d}}
\end{figure}

\begin{figure}
\centerline{\psfig{figure=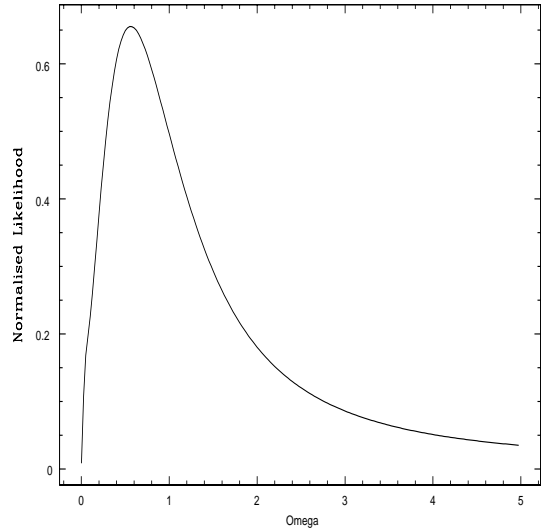,height=3.0in,width=3.0in,angle=-90}}
\caption{The 1-D marginal likelihood curve for $\Omega$. \label{fig:omega2}}
\end{figure}

In Fig.~\ref{fig:3d}
the likelihood function obtained from fitting the model $C_l$ spectra to the
data of Table~1 is shown plotted as a function of the amplitude and position
of the Doppler peak. The position is parameterized via the value of $\Omega$,
assuming that the cosmological constant is zero. The highly peaked nature of the likelihood
function in Fig.~\ref{fig:3d} is good evidence for the presence of a Doppler peak
localised in both position ($\Omega$) and amplitude. In Fig.~\ref{fig:omega2}
we show the 1-D marginal likelihood curve for $\Omega$, obtained by 
marginalising the likelihood function over $C_2$ and the peak amplitude, $A_{peak}$. 
The best fit value of $\Omega$ is 0.7 with an allowed 68\%
range of $0.2 \le \Omega \le 1.5$.

In Figure~\ref{fig:fig2a} the
best fit model, represented by the solid line, is shown compared to the data
points, assuming no error in the calibration of the Saskatoon observations.
The chi-squared per degree of freedom for this model is $0.9$, implying a good
fit to the data.  The peak lies at $l=263_{-94}^{+139}$ corresponding to a
density parameter $\Omega =0.70_{-0.4}^{+1.0}$; the height of the peak is
$A_{peak}=9.0_{-2.5}^{+4.5}$. (The errors correspond to the conditional likelihood function for each of the parameters.) The dashed and dotted lines show the best fit
models ($\Omega =0.70_{-0.37}^{+0.92}$, $A_{peak}=11.0_{-4.0}^{+5.0}$ and
$\Omega =0.68_{-0.4}^{+1.2}$, $A_{peak}=6.5_{-2.0}^{+3.5}$ respectively)
assuming that the Saskatoon observations lie at the upper and lower end of the
permitted range in calibration error.  

These likelihood results using the analytic form for the $C_l$ and the results from a chi-squared goodness of fit
analysis using exact models (see below) imply that independent of
calibration uncertainties in the data, current CMB data are inconsistent with
cosmological models with $\Omega \simlt 0.2$.

\begin{figure}
\centerline{\psfig{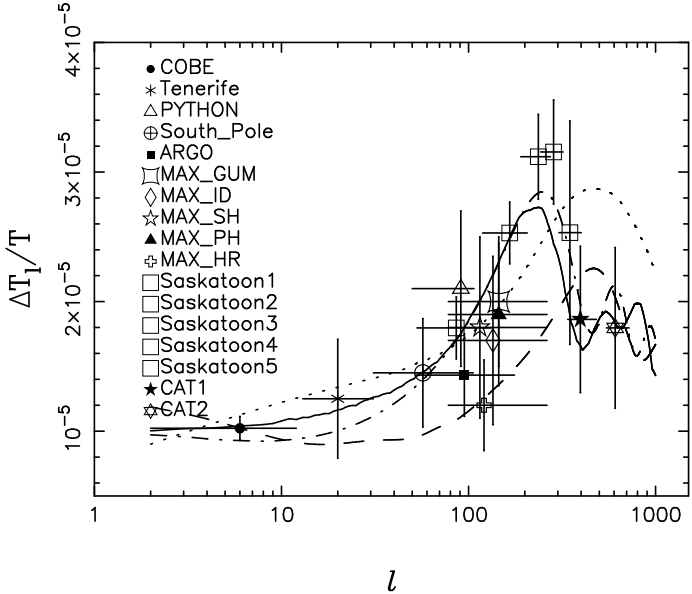}}
\caption{The data points from Table~1 compared to the exact forms of the $C_l$
for an $\Omega=1$, $\Omega_b=0.10$, $\ho=45 \kmspmpc$
standard CDM model (bold line), an $\Omega=0.3$, $\Omega_b=0.03$, $\ho=
50\kmspmpc$ open model (dashed line), a flat $\Omega = 0.3$,
$\Omega_{\Lambda}=0.7$, $\Omega_b=0.05$, $\ho=50 \kmspmpc$ model (dot-dash
line) and an example cosmic string model (dotted line) \protect\cite{strings} 
\label{fig:fig2b}}
\end{figure}

The analytic approximation to the true $C_l$ such as we use here,
is a useful general tool, but as a detailed check we have also applied the
chi-squared goodness of fit test to actual COBE normalised $C_l$ models.
In Fig.~\ref{fig:fig2b}
the data are compared to exact forms of the $C_l$ for a standard flat CDM
model, an open CDM model, a $\Lambda$ dominated model and a cosmic string model
\cite{strings}.  Allowing the model normalisation to vary within the two sigma
COBE limit we find that the standard CDM model, non-zero $\Lambda$ model and
the string model all offer acceptable ($P(\chi^2)\ge 0.05$) chi-squared fits,
whilst the probability of the open model fitting is $P(\chi^2)<0.01$.  Note
that the result for the open model ($\Omega=0.3$) using the true $C_l$ differs
from the corresponding result for the analytic form of the $C_l$: whilst in the
latter case this model is still in the allowed range of the marginal
distribution of $\Omega$, in the former case it is already excluded by the
data.  Considering a range of CDM models with varying $\Omega$, in order to
find the lowest $\Omega$ compatible with the observations, we have considered
exact models with $\ho=50 \kmspmpc$ and $\Omega_b=0.03$ for $\Omega =0.1 - 0.5$
\cite{kami}.  We find that $\Omega=0.5$ is allowed, $\Omega=0.3$ and below are
completely ruled out (95\% confidence) and $\Omega=0.4$ is excluded unless all
the Saskatoon points have the minimum allowed calibration.

We have also considered a more complete set of open models, for which partial
results can be given here. (Results over a full set of parameters will be given
in Rocha \ea, in preparation). The grid considered has $h$ values of 0.3, 0.5,
0.6, 0.7 and 0.8 (where $h=\ho/(100\kmspmpc)$), and baryon density 
$\Omega_b$ values of 0.01,
0.03, 0.06, together with $\Omega_b=0.0125h^{-2}$ and $0.024h^{-2}$ for each of
the above values of $h$.  The $\Omega$ range considered is 0.1 to 1.0 in steps
of 0.1. (These models were kindly provided by N. Sugiyama).  The unavailability
of exact models for $\Omega>1$ limits some of the statistical conclusions we can
draw here, but the results are still of interest. Assuming case (i) for the
calibration and allowing the model normalisation to vary within the two sigma
COBE limit we find that the best fit model has $\Omega =0.7$, $\ho=50 \kmspmpc$
and $\Omega_b=0.096$. This best fit value of $\Omega$ gives good agreement with
the results obtained using the analytic approximation. Marginalizing over the
other parameters, we obtain an allowed 68\% range for $\Omega$ of $0.5\le
\Omega \le 1.0$. (The upper limit of 1 is due to the cutoff in the range of
models considered.)

The situation for models in which structure formation is initiated by cosmic
strings \cite{strings,pen} is now more complex, since some of the predictions
for the power spectra for strings have recently changed.  Previous calculations
for the cosmic strings model \cite{strings} and low $\Omega$ CDM models both
have the first Doppler peak occurring in roughly the same position in $l$, so
it might be thought surprising that only the latter are eliminated by the
current data. This is traceable to the form of the low $\Omega$ power spectra
in the range $l\simeq 10$--100, where in order to match the COBE normalization
at low $l$, the models are forced to have values which are too low compared to
the data over the intermediate angular scale range. It is likely, however, that
these string models {\em will\/} be eliminated if more accurate data on the CAT
range of angular scales confirms the existing CAT results \cite{cat}.  More
recent calculations of topological defect theories indicate that the Doppler
Peak is strongly suppressed \cite{pen} and these predictions are likely to be
ruled out by the current data.

\begin{figure}
\centerline{\psfig{figure=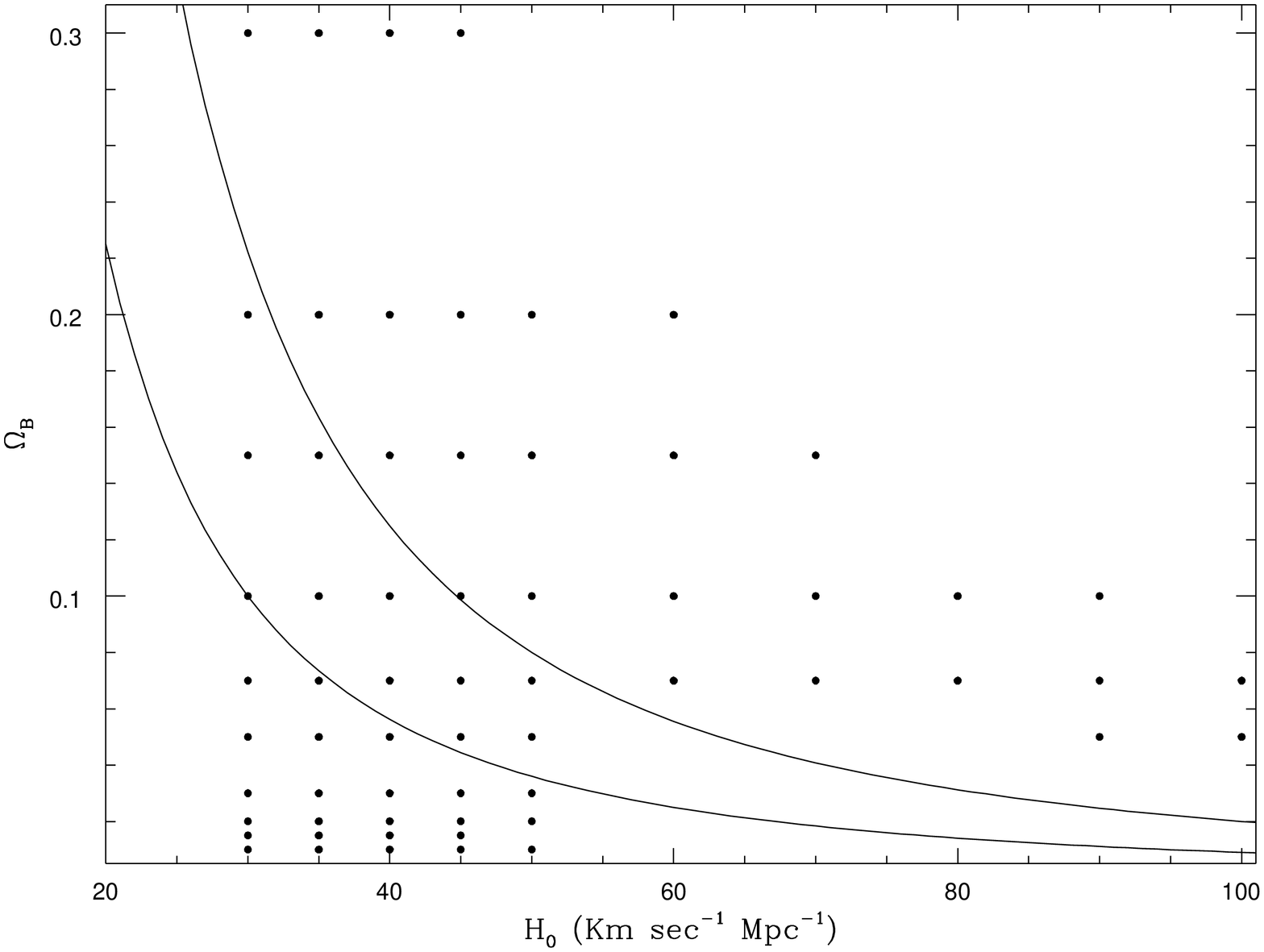,height=3.0in,width=3.0in,angle=-90}}
\caption{The COBE normalised CDM models with acceptable ($P(\chi^2) \ge 0.05$)
chi-squared fits to the CMB data (assuming the nominal Saskatoon calibration)
are plotted as dots in the $\Omega_b$-$\ho$ space. Overlying the constraint of
$0.009 \le \Omega_b h^2 \le 0.02$ \protect \cite{copi} imposed by
nucleosynthesis gives the allowed models lying between the solid curves. Models
with $\ho > 50 \kmspmpc$ are not allowed by the combined
constraint. \label{fig:omh2}}
\end{figure}

In order to set constraints on the Hubble constant in a flat universe, in
Fig.~\ref{fig:omh2}
we have considered COBE normalised standard CDM models (provided by
G. Efstathiou) for a range of $\Omega_b$ and $\mbox{H}_0$. All these models
have $\Omega=1$, $\Lambda=0$, $n=1$ and zero tensor (gravitational wave)
component. They are thus somewhat specialized, and it is well-known that they do
not provide good fits for the matter power spectrum on smaller scales. 
However, we believe our results for $\ho$ are still of interest in indicating
the type of constraints that will be available in the future, when the
increased quality of the CMB data will allow more parameters to be fitted
simultaneously.
Our method is as follows:
a dot is placed in the appropriate place in
the parameter space if the exact power spectrum corresponding to these
parameters gives a fit to the data in Table~1 with an acceptable $\chi^2$ value
($P(\chi^2) \ge 0.05$). A blank is left at that position if not.  Overlying
these power spectrum constraints is the limit $0.009 \simlt \Omega_b h^2 \simlt
0.02$ provided by nucleosynthesis of the light elements \cite{copi}. As shown,
the models offering an acceptable chi-squared fit to the CMB power spectrum,
whilst simultaneously satisfying nucleosynthesis constraints, encompass
$0.05\le \Omega_b \le 0.2$, $30\kmspmpc \le \ho \le 50 \kmspmpc$.  Allowing for
the lowest Saskatoon data calibration relaxes the constraints up to $\ho=70
\kmspmpc$. In general, recent optical and Sunyaev-Zel'dovich observations of
the Hubble constant \cite{ho3,ho1,ho2,lasenbyandjones} imply \ho in the range
$50-80 \kmspmpc$.
Since this current paper was first submitted, a recent alternative comparison of CMB data with models \cite{line} has appeared, which supports our conclusions that low values of \ho are favoured by the current CMB data.
 
Fixing $\ho=50 \kmspmpc$ and fitting for the spectral index
$n$ of the primordial fluctuations we find $n=1.1 \pm 0.1$ (68 \%
confidence). For these models, in the case of power law inflation
\cite{liddle}, this tight limit rules out a significant gravity wave
background, but agrees well with the prediction of $n\simeq 1.0$ for scalar
fluctuations generated by inflation.  

We also considered a set of tilted flat CDM models using the Seljak and
Zaldarriaga CMB code \cite{Seljak} and computed the marginal and
conditional distributions of the parameters. Results over a full set of
parameters will be given in Rocha \ea\ (in preparation), but two sample results will
 be given here to indicate the typical constraints that emerge.
Considering the nominal Saskatoon calibration case  with
superimposed BBN constraints,
we find a best fit model with $H_{0}=30 \kmspmpc$,
$\Omega_{b}=0.22$, $n=0.92$ and $Q_{rms-ps}=17.95$ \uk.
The marginal distributions of $H_{0}$ and $n$ give 68\% confidence
intervals of 
$30 \kmspmpc \le H_{0} \le 55 \kmspmpc$ and $0.85 \le n \le 1.18$ 
 respectively.

\section{Conclusions}

Our current results provide good
evidence for the Doppler peak, verifying a crucial prediction of cosmological
models and providing an interesting new measurement of fundamental cosmological
parameters. In Rocha \ea\ \shortcite{gr},
a detailed comparison of the CMB data is made with the theoretical power
spectra predicted by a range of flat, tilted, reionized, open models and models
with non-zero cosmological constant.
The existence of the Doppler peak has important consequences for
the future of CMB astronomy, implying that our basic theory is correct and that
improving our constraints on cosmological parameters is simply a matter of
improved instrumental sensitivity and ability to separate out foregrounds.  New
instruments such as VSA \cite{lasenbyandhancock}, MAP and the proposed
Planck Surveyor satellite \cite{cobras} will provide this improved sensitivity and
should delimit $\Omega$ and other parameters with unprecedented precision.

\section*{Acknowledgements}

We wish to thank all the members of the CAT and Tenerife teams for their help
and assistance in this work.  We thank G. Efstathiou and N. Sugiyama for access
to their theoretical power spectra and B. Netterfield for supplying the
Saskatoon window functions.  S. Hancock wishes to acknowledge a Research
Fellowship at St.\ John's College, Cambridge, U.K.  G. Rocha wishes to
acknowledge a JNICT Studentship from Portugal and a NSF grant EPS-9550487 with matching support from the state of Kansas and
from a K$^*$STAR First award.

\label{lastpage}


\begin{thebibliography}{}
\bibitem[\protect\citename{Bennett \ea\ }1996]{bennett}
        Bennett C.L. \ea, 1996, ApJ., 464, L1
\bibitem[\protect\citename{Bond }1995a]{bond1} 
	Bond J.R., 1995a, ``{\em Cosmology and Large Scale Structure}''
	ed. Schaeffer, R.
	Elsevier Science Publishers, Netherlands, Proc. Les Houches School, Session LX,
	August 1993
\bibitem[\protect\citename{Bond }1995b]{bond2}
	Bond, J.R., Astrophys. Lett. and Comm., 1995b, 32, 63
\bibitem[\protect\citename{Bond \& Eftstathiou }1987]{be87}
	Bond, J.R., Efstathiou, G.P., 1987, MNRAS, 226, 655
\bibitem[\protect\citename{Cheng \ea\ }1994]{msam1} 
	Cheng E.S. \ea, 1994, ApJ., 422, L37
\bibitem[\protect\citename{Cheng \ea\ }1996]{msam2} 
	Cheng E.S. \ea, 1996, ApJ., 456, L71
\bibitem[\protect\citename{Copi, Schramm \& Turner }1995]{copi} 
	Copi C.J., Schramm D.N., Turner M.S., 1995, Science, 267, 192
\bibitem[\protect\citename{Crittenden \ea\ }1993]{crittenden}
	Crittenden, R., Bond, J.R., Davis, R.L., Efstathiou, G.,
	Steinhardt, P.J., 1993, Phys. Rev. Lett., 71, 324
\bibitem[\protect\citename{Davis \ea\ }1992]{cdm1}
	Davis, M., Efstathiou, G., Frenk, C.S., White, S.D.M, 1992, Nature,
	356, 489
\bibitem[\protect\citename{De Bernardis \ea\ }1994]{argo} 
	De Bernardis P. \ea, 1994, ApJ., 422, L33
\bibitem[\protect\citename{Efstathiou }1989]{cdm2} 
        Efstathiou, G.P., 1989, in
	``{\em Physics of the Early Universe}'', proceedings of the
	thirty-sixth Scottish Universities Summer School in physics 1989,
	p361, eds. Peacock, J.A., Heavens, A.F., Davies, A.T. 
\bibitem[\protect\citename{Fischer \ea\ }1995]{fischer} 
	Fischer, M.L. \ea, 1995, ApJ., 444, 226
\bibitem[\protect\citename{Freedman \ea\ }1994]{ho1}
	Freedman, W.L. \ea, 1994, Nature, 371, 757
\bibitem[\protect\citename{Gundersen \ea\ }1995]{spole} 
	Gundersen, J.O \ea, 1995, ApJ., 443 L57
\bibitem[\protect\citename{Guth }1981]{guth} 
	Guth, A.H., 1981, Phys. Rev. D, 23, 347
\bibitem[\protect\citename{Hancock \ea\ }in press]{me96}
	Hancock, S., Gutierrez, C.M., Davies, R.D., 
	Lasenby, A.N., Rocha, G., Rebolo, R., Watson, R.A.,
	and Tegmark M., 1997, MNRAS, in press
\bibitem[\protect\citename{Hancock \ea\ }1994]{nature94} 
	Hancock, S., Davies, R.D., Lasenby, A.N., De La Cruz, C.M.G., 
	Watson, R.A., Rebolo, R., Beckman, J.E., 1994, Nature, 367, 333
\bibitem[\protect\citename{Hu \& Sugiyama }1995]{hu} 
	Hu W., Sugiyama N., 1995, ApJ., 444, 489
\bibitem[\protect\citename{Kamionkowsky \ea\ }1994a]{kamio}  
	Kamionkowsky, M., Spergel, D.N., Sugiyama, N.,
	1994a, ApJ., 426, L57
\bibitem[\protect\citename{Kamionkowsky \ea\ }1994b]{kami} 
	Kamionkowsky, M., Ratra, B., Spergel, D.N., and
	Sugiyama, N., 1994b, ApJ., 434, L1	
\bibitem[\protect\citename{Kennicutt, Freedman \& Mould }1995]{ho2}
	Kennicutt, R.C., Freedman, W.L., Mould, J.R., 1995,
	ApJ., 110, 1476 
\bibitem[\protect\citename{Lasenby \& Hancock }1995]{lasenbyandhancock}
	Lasenby, A.N., Hancock, S., 1995, Proc. of : {\em ``Current Topics 
	in Astrofundamental Physics: The Early Universe''},
	p327,  eds. Sanchez, N., Zichichi, A., Kluwer
\bibitem[\protect\citename{Lasenby \& Jones }1997]{lasenbyandjones}
	Lasenby, A.N., Jones M.E., 1997, Proc. of : {\em ``The 
	Extragalactic distance Scale''},
	p76, eds Livio N., Donahue M., Panagia N., CUP
\bibitem[\protect\citename{Liddle }1997]{liddle97} 
	Liddle, A.R., 1997, Proceedings of `From Quantum Fluctuations 
	to Cosmological Structures', Casablanca,
	Morocco, December 1996, in press (astro-ph/9612093)
\bibitem[\protect\citename{Liddle \& Lyth }1992]{liddle} 
	Liddle, A.R., Lyth, D.H., 1992, Phys. Letters, B291, 391
\bibitem[\protect\citename{Lineweaver \& Barbosa }in press]{line}
	Lineweaver, C.H., Barbosa, D., in press (astro-ph/9612146)
\bibitem[\protect\citename{Magueijo \ea\ }1996]{strings}
	Magueijo, J., Albrecht, A., Coulson, D., Ferreira, P.,
	1996, Phys. Rev. Lett., 76, 2617 
\bibitem[\protect\citename{Mandolesi \ea\ }1995]{cobras}
	Mandolesi, N. \ea, 1995, Planetary and Space Science, 43, 1459
\bibitem[\protect\citename{Netterfield \ea\ }1997]{sask} 
	Netterfield, C.B., Devlin, M.J., Jarosik, N., Page L.,
	and Wollack, E.J., 1997, ApJ., 474, 47 
\bibitem[\protect\citename{Ostriker \& Steinhardt }1995]{os} 
	Ostriker, J.P., Steinhardt, P.J., 1995, Nature, 377, 600
\bibitem[\protect\citename{Pen, Seljak \& Turok }1997]{pen}
	Pen, U.-L., Seljak, U., Turok, N. 1997, preprint (astro-ph/9704165)
\bibitem[\protect\citename{Pierce \ea\ }1994]{ho3} 
	Pierce, M.J., Welch, D.L., Mcclure, R.D., Van Den Bergh, S., 
	Racine, R., Stetson, P.B., 1994, Nature, 371, 385
\bibitem[\protect\citename{Ratra \ea\ }1997]{rhatra} 
	Ratra, B., Sugiyama, N., Banday, A.J., and
	Gorsky, K.M., 1997, Apj., 481 
\bibitem[\protect\citename{Rocha \ea\ }in preparation]{gr}
	Rocha, G., Hancock, S., Lasenby, A.N., Gutierrez, C.M. in preparation.
\bibitem[\protect\citename{Ruhl \ea\ }1995]{python} 
	Ruhl, J.E., Dragovan, M., Platt, S.R., Kovac, J., Novak, G.,
	 1995, ApJ., 453, L1
\bibitem[\protect\citename{Scaramella \& Vittorio }1990]{cosvariance2} 
	Scaramella R., Vittorio N., 1990, ApJ., 353, 372
\bibitem[\protect\citename{Scaramella \& Vittorio }1993]{cosvariance1} 
	Scaramella R., Vittorio N., 1993, ApJ., 411, 1
\bibitem[\protect\citename{Scott, Srednicki and White }1994]{samvariance} 
	Scott D., Srednicki M., White M., 1994, ApJ., 241, L5
\bibitem[\protect\citename{Scott, Silk \& White }1995]{scott} 
	Scott D., Silk, J., White, M., 1995, Science, 268, 5212
\bibitem[\protect\citename{Scott \ea\ }1996]{cat} 
	Scott, P.F. \ea, 1996, ApJ., 461, L1
\bibitem[\protect\citename{Seljak \& Zaldarriaga }1996]{Seljak} 
        Seljak, U., Zaldarriaga, M., 1996, ApJ, 469, 437
\bibitem[\protect\citename{Smoot \ea\ }1992]{smoot} 
	Smoot, G.F. \ea, 1992, ApJ., 396, L1
\bibitem[\protect\citename{Steinhardt }1993]{steinhardt} 
	Steinhardt, P., 1993,
	{\em Proc. of the Yamada Conference XXXVII ``Evolution of the Universe
	and its Observational Quest''}, p159, ed. Sato, K. 
\bibitem[\protect\citename{Tanaka \ea\ }1996]{max} 
	Tanaka, S.T. \ea, 1996, ApJ., 468, L81
\bibitem[\protect\citename{White \& Srednicki }1995]{whitewindow2} 
	White M., Srednicki M., 1995, ApJ., 443, 6
\bibitem[\protect\citename{White, Krauss \& Silk }1993]{whitewindow1} 
	White M., Krauss L.M., Silk J., 1993, ApJ., 418, 535
\bibitem[\protect\citename{White, Scott \& Silk }1994]{white}
	White, M., Scott, D., Silk, J., 1994, Ann. Rev. Astron. Astrophys., 32,
	319
\end{thebibliography}
\end{document}